\documentclass[12pt]{article}

\usepackage{amsmath,amssymb}
\usepackage{graphicx}

\makeatletter

\@addtoreset{equation}{section}
\makeatother

\newcommand{\Tr}{\mathop{\rm Tr}}
\newcommand{\p}{\partial}
\newcommand{\F}{\cal{F}}
\newcommand{\W}{{\cal W}}
\newcommand{\N}{{\cal N}}
\newcommand{\M}{{\cal M}}
\newcommand{\G}{{\cal G}}

\def\({\left(}
\def\){\right)}
\def\cO{{\mathcal O}}

\def\XXint#1#2#3{{\setbox0=\hbox{$#1{#2#3}{\int}$} 
\vcenter{\hbox{$#2#3$}}\kern-.5\wd0}}

\begin{document}

\setlength{\oddsidemargin}{0cm}
\setlength{\baselineskip}{7mm}

\begin{titlepage}
\renewcommand{\thefootnote}{\fnsymbol{footnote}}

\begin{flushright}
\begin{tabular}{l}
RIKEN-TH-180\\
KUNS-2244
\end{tabular}
\end{flushright}

\vspace*{3cm}
    \begin{Large}
    \begin{bf}
       \begin{center}
         {Genus-one correction to 
asymptotically free \\
Seiberg-Witten prepotential from \\
Dijkgraaf-Vafa matrix model } 
       \end{center}
    \end{bf}   
    \end{Large}
\vspace{1cm}

\begin{center}
{ \sc Mitsutoshi Fujita}\footnote    
{e-mail address : {\tt mfujita@gauge.scphys.kyoto-u.ac.jp}} 
{ \sc Yasuyuki Hatsuda }\footnote    
{e-mail address : {\tt hatsuda@riken.jp}} 
{ \sc Ta-Sheng Tai}\footnote    
{e-mail address : {\tt tasheng@riken.jp}} 
\end{center}

      \vspace{1cm}

\begin{center}
{\it Department of Physics, Kyoto University, Kyoto 606-8502, JAPAN}\\
{\it   Theoretical Physics Laboratory, RIKEN,
                    Wako, Saitama 351-0198, JAPAN}
\end{center}

\vspace{1cm}

\begin{abstract}
\noindent
We find perfect agreements on the genus-one correction to the 
prepotential of $SU(2)$ Seiberg-Witten theory with $N_f=2$, $3$ 
between field theoretical 
and Dijkgraaf-Vafa-Penner type matrix model results. 
\end{abstract}
\vfill 

\end{titlepage}
\vfil\eject

\setcounter{footnote}{0}

\section{Introduction }
Recently, 
owing to a milestone discovery made by 
Alday, Gaiotto and Tachikawa 
\cite{Alday:2009aq}, there have been 
lots of publications and research related to their work 
\cite{ Alba:2009ya}-\cite{Wyllard:2009hg}. 
In particular, Dijkgraaf and Vafa 
\cite{Dijkgraaf:2009pc} proposed a Penner type 
matrix model whose classical spectral curve can reproduce the so-called 
Gaiotto curve ${\cal{G}}$ \cite{Gaiotto:2009we}. 
Note that $\G$ consists of a punctured Riemann surface $C_{g,n}$ whose moduli space $\M_{g,n}$ 
($g$: genus, $n$: puncture) is referred to as a Teichmuller space. 
Surprisingly, $\M_{g,n}$ boils down to the space of exactly marginal 
gauge couplings of a large family of 4D $\N=2$ superconformal gauge theories 
whose weakly-coupled {\it cusps} correspond to various patterns of colliding punctures on $C_{g,n}$. 
In addition, when $(g,n)=(0,6)$ there appear {\it generalized} quiver SCFTs 
in contrast to known 
linear quiver SCFTs. 
Further studies towards this newly proposed matrix model can be found in 
\cite{Itoyama:2009sc, 
Eguchi:2009gf, Schiappa:2009cc, Mironov:2009ib}. 
Because $\G$ is a rewritten Seiberg-Witten curve which 
emerges by taking a thermodynamic limit 
of 
Nekrasov's partition function 
$Z_{\mathrm{ Nekrasov }}
=Z_{\mathrm{classical}}Z_{1\text{-}\mathrm{loop}}Z_{\mathrm{inst}}$ \cite{N1,N2,N3}, 
attempts 
towards proving an equivalence between both sides are 
naturally expected.

At the level 
of ${\F}_0$ 
(tree-level free energy), Eguchi and Maruyoshi \cite{Eguchi:2009gf} showed that 
${\F}_0$ (including asymptotically free cases) 
coincides with the original Seiberg-Witten 
prepotential \cite{Seiberg:1994rs}. 
Moreover, in \cite{Schiappa:2009cc, Mironov:2009ib} 
all-genus proofs in certain 
restricted cases are presented by executing 
exact matrix integrals and comparing them 
with $Z_{\mathrm{ Nekrasov }}$. 
Motivated by these works, in this letter 
we would like to show agreements between 
matrix model and field theoretical results on 
the genus-one free energy ${\F}_1$ of 
$\N=2$ $SU(2)$ Seiberg-Witten theory with $N_f=2$, $3$. As a matter of fact, 
we have closely followed previous approaches in \cite{Klemm:2002pa,Dijkgraaf:2002yn}.
 
In Section 2, we begin with 
a topologically twisted theory living on a hyperK{\"a}hler 
manifold and extract a physical ${\F}_1$. In Section 3, 
a matrix model proposed by \cite{Eguchi:2009gf} is used to 
compute ${\F}_1$. We summarize our result in Section 4.

\section{Field theory}

Gravitational couplings of the form 
$\int d^4x {\F}_g R^2_+ F_+^{2g-2}$ 
$(g\ge 1)$ due to 
a curved four-manifold $\M_4$ 
give rise to a corrected Seiberg-Witten 
prepotential in terms of a {\it genus} expansion: 
\begin{eqnarray}
&&{\cal F}=\sum_{g\ge 0} \hbar^{2g-2} {\cal{F}}_g (
\boldsymbol{a},\boldsymbol{m})=-\log Z_{\mathrm{ Nekrasov }}
 ,\nonumber\\ 
&&\boldsymbol{a}: \mathrm{Coulomb ~branch ~parameters }, ~~~~~~~
\boldsymbol{m}: \mathrm{hypermultiplet ~masses}.
\end{eqnarray}
Here, $R_+$ and $F_+=\hbar$ are 
the self-dual part of the Riemann curvature 
and the graviphoton field strength respectively. 
In particular, when $\M_4$ is Euclidean, the genus-one 
correction is given by 
\begin{eqnarray}
\label{KK1}
&&\int d^4x {\cal{F}}_1  \Tr R^2_+ 
=\frac{1}{2}{\cal{F}}_1 (\chi - \frac{3}{2}\sigma),~~~~~~~R_{\pm}=\frac{1}{2}(R\pm 
R^{\ast})\nonumber\\ 
&&\chi=\frac{1}{32\pi^2}\int R 
\wedge R^{\ast},~~~~~~~~~~~
\sigma=\frac{1}{24\pi^2}\int R
\wedge R
\end{eqnarray}
where $\chi(\M_4)$ and $\sigma(\M_4)$ denote the 
Euler number and the Hirzebruch signature respectively. 

Now, let us focus on a topologically twisted $\N=2$ $SU(2)$ theory with 
hypermultiplets living 
on $M_4$. The low-energy partition function looks like 
\begin{eqnarray}
\label{}
&&Z=\int [du]A^{\chi} B^{\sigma} \exp{\big(-S \big)},\nonumber\\ 
&&A=\alpha \sqrt{\frac{\p u}{\p a}}, ~~~~~~~
B=\beta \Delta_{SW}^{\frac{1}{8}}, ~~~~~~~
\alpha, ~\beta: \mathrm{constants}\nonumber
\end{eqnarray}
where $u$ stands for the gauge- and monodromy-invariant coordinate of the complex  one-dimensional Coulomb branch. 
Forms of $A$ and $B$ appearing above are required to ensure 
the modular invariance of $Z$ and necessarily 
cancel the modular anomaly caused by $[du]$ 
\cite{Witten:1995gf,Moore:1997pc}. 
These considerations then define 
a field theoretical version of the coupling to gravity, i.e. 
\begin{eqnarray}
\label{KK}
A^{\chi} B^{\sigma}=\exp \Big( b(u) \chi + c(u) \sigma \Big), ~~~~~~~
b(u)=\frac{1}{2}\log \Big(\frac{du}{da}\Big), 
~~~~~~~c(u)=\frac{1}{8}\log \Big(\Delta_{SW}\Big).
\end{eqnarray}
Here, $a$ is the electric period integral of 
the corresponding Seiberg-Witten curve and 
$\Delta_{SW}$ denotes its discriminant. 
When $M_4$ is hyperK{\"a}hler ($\sigma=-2\chi/3$) 
or a $K3$ manifold ($\chi=24$ and $\sigma=-16$), the effect of 
twist%
\footnote{The topological twist is 
performed through replacing 
$SU(2)_+ \subset SO(4) \cong SU(2)_+ \times SU(2)_-$ 
by the 
diagonal part of $SU(2)_+ \times SU(2)_R$ where 
$SU(2)_R$ represents the $R$-symmetry. For 
hyperK{\"a}hler manifolds, that 
no holonomy is involved in 
$SU(2)_+$ implies that to twist will not be visible. } 
is not visible and \eqref{KK} of a 
twisted theory becomes 
compatible with that of 
a physical theory. Therefore, equating \eqref{KK1} with 
$\big( b(u) \chi + c(u) \sigma \big)$  of a hyperK{\"a}hler $M_4$ in \eqref{KK} we see that  
\begin{eqnarray}
\label{FR}
{\F}_1= b(u) -\frac{2}{3} c(u) . 
\end{eqnarray}
In order to determine $b(u)$ and $c(u)$, one needs an explicit 
Seiberg-Witten curve $\Sigma$ 
\begin{eqnarray}
\label{MSW}
\Sigma: \prod_{I=0}^n (t-t_I) v^2   = M_{n+1}(t) v+  U_{n+1}(t), ~~~~~~~
(t,v)\in(\mathbb{C}^{\ast}-\{ t_0, \cdots, t_n \}) \times \mathbb{C}
\end{eqnarray}
and notices that 
($\lambda_{SW}$: Seiberg-Witten one-form) 
\begin{eqnarray}
\frac{d a}{d u} = \frac{d }{d u}\oint_{A} \lambda_{SW}.
\nonumber
\end{eqnarray}
The subscript of polynomials $M$ and $U$ denotes their degree. 
According to \cite{Witten:1997sc}, 
$\Sigma$ arises from an M-theory lift 
of Type IIA D4- and NS5-branes engineering 
$\N=2$ $SU(n+1)$ Yang-Mills theory 
with fundamental matters which are encoded at two asymptotic ends $(t=0, \infty)$ of $\Sigma$. 
The gauge coupling $\tau_I$ of $I$-th gauge factor of a conventional linear quiver is expressed in terms of 
$t=\exp \big( x^6 + i\frac{x^{10}}{R_{M}} \big)$ 
($R_M$: M-circle radius) parameterizing a cylinder along $(x^6, x^{10})$: 
\begin{eqnarray}
i\pi \tau_I =\log \frac{t_{I-1}}{t_I}, ~~~~~~~
~~~~~
\tau=\frac{\theta}{\pi} + 
\frac{8\pi i}{g^2}.
\nonumber
\end{eqnarray}
As shown in \cite{Gaiotto:2009we},
through performing a change of variables $v=xt$ and 
certain proper M{\"o}bius transformation on $t$, one finally obtains a so-called Gaiotto curve:
\begin{eqnarray}
\label{MSW2}
{\G}: x^2= \phi_2^{SW} (z), ~~~~~~~
{xdz}\equiv\lambda_{SW}.
\end{eqnarray}

For the simplest $SU(2)$ Seiberg-Witten theory with $N_f=2$ and $3$, 
the period integral $a(u)$ had been computed by Ohta \cite{Ohta:1996fr} in terms 
of a large-$u$ expansion (weak coupling expansion). 
Therefore, it is straightforward to evaluate $b(u)$ for $N_f=2$: 
\begin{align} 
b(u)&=-\frac{1}{2}\log\( \frac{da}{du} \) \notag \\
&=\frac{3}{4}\log 2-\frac{1}{4} \log \zeta-\frac{3\Lambda^2}{2048}(\Lambda^2+64m_1m_2)\zeta^2
+\frac{15\Lambda^4}{2048}(m_1^2+m_2^2)\zeta^3+\cO(\zeta^4)
\label{a1}
\end{align}
where $\zeta=1/u$. Similarly, for $N_f=3$, 
\begin{align}
b(u)
&=\frac{1}{2}\log 2-\frac{1}{4} \log \zeta-\frac{\Lambda^2}{2048}\zeta \label{a2}\\
&\hspace{0.5cm}-\frac{\Lambda}{8388608}\Big( 7\Lambda^3+12288(m_1^2+m_2^2+m_3^2)\Lambda+786432m_1m_2m_3  \Big)\zeta^2
+\cO(\zeta^3).\notag 
\end{align}
We have denoted flavor bare masses and 
the dynamical scale by $m_i$'s and $\Lambda$ respectively. 
In Section 3, we will find perfect agreements with these results 
in carrying out a computation via the 
matrix model proposed by Eguchi and Maruyoshi \cite{Eguchi:2009gf}.

\section{Matrix model}
Before computing the genus-one free energy ${\F}_1$, 
we first give a brief introduction about the newly proposed Dijkgraaf-Vafa 
matrix model. 
Without the background charge $Q=b +b^{-1}$ $(b=i)$, 
in computing correlators of vertex operators 
$\langle  \prod_{i} V_i (\xi_i) \rangle$ in Liouville 
theory, Dijkgraaf and 
Vafa 
\cite{Dijkgraaf:2009pc} 
have replaced the usual 
Liouville wall by a {\it chiral} one $\int d\xi  e^{\sqrt{2}b\phi(\xi)}$. 
This results in a 
{\it hermitian} matrix model with an usual 
Vandermonde, and inserted operators $V_{i}(\xi_i)=e^{i\sqrt{2} p_i\phi (\xi_i)}$ as a whole consequently 
lead to a 
logarithmic potential of Penner type, i.e. 
\begin{eqnarray}
&&Z_{\mathrm{ DV }}=\Big\langle  \prod_{i} V_i (\xi_i)
\Big\rangle_{\mathrm{chiral ~Liouville}}=\int_{N\times N} dM \exp \Big( 
\frac{1}{g_s} \Tr{\cal{W}} (M)\Big) =\exp \Big(-\sum_{g\ge 0} g_s^{2g-2} {\cal{F}}_g \Big),\nonumber\\
&&{\cal{W}} (M)=\sum_{i} \mu_i \log(M-\xi_i), 
~~~~~~~\mu_i=2g_s p_i, ~~~~~~~\sum_{i} \mu_i+\mu_0  + 2g_s N=0,\nonumber\\
&&p_i,~ N\to \infty, ~~~~~~~
g_s\to 0, ~~~~~~~
\mu_i, ~g_s N=\mathrm{fixed}.
\label{PPP}
\end{eqnarray}
The charge conservation is 
respected by placing $\mu_0$ units at infinity. 

Interpreting the above chiral free boson $\phi$ as a Kodaira-Spencer (collective) field which is especially powerful in dealing with 
quantizing the Riemann surface complex moduli, 
one can express the matrix model $quantum$ 
spectral curve as 
\begin{eqnarray}
\label{1}
&&-i\Big\langle \p \phi(z) 
\Big\rangle=\Big( -\nu_0 z^{-1}+\sum_{n>0} n \nu_n z^{n-1} 
+g_s^2 \sum_{n\ge 0} z^{-n-1}\frac{\p}{\p \nu_n} \Big)Z_{\mathrm{ DV }}\nonumber
\end{eqnarray}
with $z$ parameterizing it. $\nu_n$ and its conjugate are referred to as symplectic 
{\it coordinates} of the moduli space. Eventually, 
$\langle \p \phi(z) \rangle_{g_s \to 0}$ just 
reduces to a Gaiotto curve $\G$ of $\N=2$ 
$SU(2)$ SCFTs as will be explained more below. 
Dijkgraaf and 
Vafa's intuition seems due to the marvelous discovery of 
Alday, Gaiotto and Tachikawa \cite{Alday:2009aq} 
relating correlators in Liouville theory to 
Nekrasov's partition function. Recall that 
$\G$ was yielded by reorganizing a Seiberg-Witten 
curve which emerges via taking a thermodynamic limit 
($\hbar \to 0$) of 
$Z_{\mathrm{ Nekrasov }}$. 
It is thus very tempting to recognize a full equivalence 
$Z_{\mathrm{ DV }}=Z_{\mathrm{chiral ~Liouville~correlator}}
=Z_{\mathrm{ Nekrasov }}$ with $g_s=\hbar$. This line has been 
pursuit in \cite{ Eguchi:2009gf, Schiappa:2009cc, Mironov:2009ib}%
\footnote{An early attempt towards interpreting Nekrasov's 
partition function as a kind of tachyon's scattering amplitude 
in the self-dual 
$c=1$ string theory can be found in 
\cite{Tai:2007vc}. 
There, vertex operators made of a 
collective field of a Fermi fluid are inserted at $q$-numbered
positions on two asymptotic regions of a sphere.}.

As pointed out by AGT, one can yield 
a quadratic 
Seiberg-Witten differential from Ward identities in 
Liouville theory%
\footnote{We must apologize for using $\phi$ in expressing 
quadratic differentials and Kodaira-Spencer 
fields simultaneously.}: 
\begin{eqnarray}
&&\phi_2(z) dz^2 = \frac{\Big\langle T(z) \prod_{i} {\cal O}_i  (z_i)
\Big\rangle}{\Big\langle \prod_{i} {\cal O}_i  (z_i)
\Big\rangle}, ~~~~~~~T(z): \mathrm{stress ~tensor}, 
\nonumber\\
&&\phi_2 (z)dz^2 ~ \to ~
\phi_2^{SW} (z)dz^2=\lambda^2_{SW}, 
~~~~~~~\mathrm{when}~
1 \gg \epsilon_{1,2}.
\label{TTT}
\end{eqnarray}
Note that 
${\cal O}$'s are inserted at the level of 
conformal blocks in Liouville theory, while $\epsilon_{i}$ denotes the non-self-dual 
graviphoton field strength appearing in Nekrasov's formula. 
Through $x^2=\phi_2^{SW} (z)$ one obtains 
a Gaiotto curve which is a double cover of a punctured sphere 
with cuts. 
From \eqref{TTT}, 
a reasonable analogy is 
strongly recommended in the aforementioned $Z_{\mathrm{ DV }}$. 
Because 
a stress tensor on the hermitian matrix model side 
can be 
defined through a Kodaira-Spencer field, i.e.   
\begin{eqnarray}
T(z) = -\frac{1}{2}(\p \phi)^2=
\sum_{n\in\mathbb{Z}} L_n z^{-n-2},\nonumber
\end{eqnarray}
a classical spectral curve 
emerging in large-$N$ limit is written as  
\begin{eqnarray}
\Big\langle T(z)\Big\rangle 
=-\frac{1}{2}\Big\langle \p \phi(z)^2 
\Big\rangle ~\to ~\frac{1}{2}\W' (z)^2  +2f(z)
\label{x2}
\end{eqnarray}
where the average is w.r.t. $Z_{\mathrm{ DV }}$. 
Equivalently, 
\begin{eqnarray}
\label{}
-x=i\Big\langle\p \phi(z)\Big\rangle=-{\cal{W}}'(z) -2\omega(z), 
~~~~~~~~~~\omega(z)=g_s \Tr \Big\langle 
\frac{1}{z-M}\Big\rangle  \nonumber
\end{eqnarray}
with which an $SU(2)$ Gaiotto curve is identified by Dijkgraaf and Vafa. 
The arrow in \eqref{x2} is completely owing to 
a factorization of the resolvent operator 
at large-$N$ limit: 
\begin{eqnarray}
g_s^2 \Big\langle\Tr 
 \frac{1}{z-M} \Tr 
 \frac{1}{z-M}\Big\rangle =
\omega(z)^2 \nonumber
\end{eqnarray}
such that the all-genus loop equation becomes 
\begin{eqnarray}
\omega(z)^2 + \omega(z)\W'(z)=g_s
\Big\langle \frac{\W'(z)-\W'(M)}{z-M} \Big\rangle=f(z).
\nonumber
\end{eqnarray}
In 
\cite{Eguchi:2009gf}, 
$x^2=\W'(z)^2 + 4f(z)$ 
was shown to coincide with $x^2=\phi_2^{SW}(z)$ in 
\eqref{TTT} 
by fully exploiting known properties of standard Seiberg-Witten curves.

Let us pause to see a canonical example $SU(2)$ $N_f=4$. 
Four insertions $(V_0,V_1,V_2,V_3)$ are prescribed to be located at $(\infty,q,1,0)$ in order and $V_0$ at $\infty$ will never show up in $\W$ though. It is evident that 
residues of $x_{DV}(z)$ at $(\infty,q,1,0)$ 
correspond to momenta of $V_i$'s which are identified with 
flavor bare masses according to AGT dictionary. 
Also, $q$ stands for the cross-ratio of four distinct 
punctures on 
a sphere and hence 
lives on $\mathbb{CP}^1 \backslash \{ 0, 1, \infty \}$. More explicitly, 
one is allowed to choose certain 
M{\"o}bius transformation $f$: 
\begin{eqnarray}
f(z)=\frac{az+b}{cz+d}, ~~~~~~ad-bc\ne 0,
~~~~~~a,b,c,d\in\mathbb{C}\nonumber
\end{eqnarray}
which brings three points $(z_1, z_2, z_3)$ on a sphere 
to the triple $(0,1,\infty)$, 
while $z_4$ is mapped to $f(z_4)=q$. 
Ultimately, 
$q$ is just $q_{UV}=e^{i\pi \tau_{UV}}$ 
because this interpretation is totally supported by the known space of the 
exactly marginal (ultra-violet) gauge coupling constant $\tau_{UV}$.

\subsection{Genus-one correction} 
For four insertions at $(\infty,q,1,0)$ in \eqref{PPP}, it is obvious that 
there will be two critical points (zeros) 
for $\W'(z)=0$ of $Z_{\mathrm{ DV }}$ if one recalls 
that $V_0(\infty)$ does not show up. 
When quantum effects introduced by the resolvent are incorporated, 
they blow up into two cuts whose filling fractions 
$N_1$ and $N_2$ subject to the constraint $N_1 + N_2=N$ 
($N$: rank of $M$). 
The classical spectral curve is a double cover of a punctured 
sphere with two cuts and this kind of two-cut model has 
been 
fully investigated \cite{Ambjorn:1992gw,Akemann:1996zr}. Borrowing Akemann's analysis, we are able to have 
the genus-one free energy expressed in an universal form%
\footnote{Strictly speaking, this form was prescribed 
for a polynomial potential $\W(z)$. } 
\begin{eqnarray}
\label{F11}
{\F}_1=  -\frac{1}{24} \sum_{i=1}^4 \log M_i 
   -\frac{1}{12} \log \Delta -\frac{1}{2} \log |K(\ell)| 
   +  
\frac{1}{4} \log |(x_1-x_3)(x_2-x_4)|
\end{eqnarray}
where 
\begin{eqnarray}
&&M_i =\oint_{{{\cal C}}} \frac{dz}{2\pi i}
\frac{{{\cal W}'}(z)}{(z-x_i)
\sqrt{\prod^4_{i=1}(z-x_i)}},~~~~~~
{{\cal C}}: \mathrm{contour ~encircling ~both~cuts}
\nonumber\\
&&\ell^2=\frac{(x_1-x_4)(x_2-x_3)}{(x_1-x_3)(x_2-x_4)}, ~~~~~~
\Delta=\prod_{i<j} (x_i -x_j)^2.
\end{eqnarray}
$[x_1, x_2]$ and $[x_3,x_4]$ stand for branch points 
of these two cuts with their cross-ratio denoted by $\ell^2$, while $K(\ell)$ is the complete elliptic integral of the first kind. One can soon realize that 
$M_i=0$ when the contour ${{\cal C}}$ is deformed to enclose $\infty$. Divergent terms like 
$\log {M_i}$ will then be omitted. 
To deal with subsequent terms without knowing 
explicitly four branch points, we can appeal to 
a very helpful formula suggested by Masuda and Suzuki \cite{Masuda:1996xj}. 
That is, noting the equality between a hypergeometric function and 
a complete elliptic integral ${}_2F_1(\frac{1}{2},\frac{1}{2};1;\ell^2)=
\frac{2}{\pi}K(\ell)$, 
one is able to rewrite the last two terms 
in \eqref{F11} as 
\begin{align}
-\frac{1}{2}\log \Big( \frac{\pi}{2}(-D)^{-\frac{1}{4}} \,_2F_1 ( \frac{1}{12},\frac{5}{12};1;-\frac{27\Delta_g}{4D^3})\Big).
\nonumber
\end{align}
Here, ${}_2F_1(\alpha,\beta;\gamma;\delta)$ is the hypergeometric function, $\Delta_g$ is the discriminant of certain quartic polynomial $g(y)=y^4+ay^3+by^2+cy+d$ whose four roots are previous $( x_1, x_2, x_3,x_4)$ with 
\begin{align}
&\Delta_g =-\Big( 27a^4d^2+a^3c(4c^2-18bd)+ac(-18bc^2+80b^2d+192d^2) \nonumber \\
&\hspace{1cm}+a^2(-b^2c^2+4b^3d+6c^2d-144bd^2)+4b^3c^2+27c^4 \nonumber\\
&\hspace{1cm}-16b^4d-144bc^2d+128b^2d^2-256d^3 \Big) \nonumber
\end{align}
and $D\equiv -b^2+3ac-12d$.

For asymptotically free $SU(2)$ $N_f=2$ and $3$, the classical spectral curve can be derived from the original $N_f=4$ one via scaling limits which amount to decoupling extremely massive flavors. By adhering to \cite{Eguchi:2009gf} and adopting 
their convention, 
adequate candidates 
responsible for the aforementioned quartic $g(y)$ 
extracted from the classical spectral curve are then 
\begin{align}
R_4(y)=y^4+\frac{4M_+}{\Lambda_2}y^3+\frac{4v}{\Lambda_2^2}y^2+\frac{4\tilde{M}_+}{\Lambda_2}y+1
\label{222}
\end{align}
and 
\begin{align}
Q_4(y)&=y^4+\frac{1}{M_0^2}\( -v-M_0^2+M_2^2+\frac{1}{2}\tilde{M}_+\Lambda_3\)y^3 \nonumber \\
&\hspace{1cm}+\frac{1}{M_0^2}\(v+\frac{\Lambda_3^2}{4}-\frac{3}{2}\tilde{M}_+\Lambda_3\)y^2
+\frac{1}{M_0^2}\(-\frac{\Lambda_3^2}{2}+\tilde{M}_+ \Lambda_3\)y+\frac{\Lambda_3^2}{4M_0^2}
\label{333}
\end{align}
for $N_f=2$ and 3 respectively. 
Through the following identification in \eqref{222}:
\begin{eqnarray}
v=4u,\;\;\; M_+=2m_1,\;\;\; \tilde{M}_+=2m_2,\;\;\; \Lambda_2=\Lambda,
\end{eqnarray} 
the last two terms 
in \eqref{F11} are thus found to be ($\zeta=1/u$)
\begin{align}
-\frac{1}{2}\log\frac{\pi}{8} -\frac{1}{2}\log \Lambda-\frac{1}{4}\log \zeta -\frac{3\Lambda^2}{2048}(\Lambda^2+64m_1m_2)\zeta^2
+\frac{15\Lambda^4}{2048}(m_1^2+m_2^2)\zeta^3+\cO(\zeta^4)
\label{a3}
\end{align}
expressed in terms of a large-$u$ expansion. 
Similarly, through 
\begin{eqnarray}
v=4u,\,\;\;\; M_+=2m_1,\,\;\;\;M_-=2m_2,\,\;\;\;\tilde{M}_+=2m_3,\,\;\;\;
\Lambda_3=\frac{\Lambda}{2}
\end{eqnarray} 
in \eqref{333}, the last two terms 
in \eqref{F11} are found to be
\begin{align}
&-\frac{1}{2}\log\frac{\pi}{4}-\frac{1}{2}\log|m_1-m_2|-\frac{1}{4} \log \zeta-\frac{\Lambda^2}{2048}\zeta \label{a4}\\
&-\frac{\Lambda}{8388608}\Big( 7\Lambda^3+12288(m_1^2+m_2^2+m_3^2)\Lambda+786432m_1m_2m_3 
\Big)\zeta^2
+\cO(\zeta^3).\notag 
\end{align}
As stressed before, the 
matrix model classical spectral curve is just the same 
as the corresponding Gaiotto curve (rearranged Seiberg-Witten curve), 
henceforth we still have same discriminant 
$\Delta_{SW}=\Delta$ in 
\eqref{KK} and \eqref{F11} 
even after decoupling massive flavors%
\footnote{In fact, this can be easily checked by 
comparing our above $\Delta_g$ with the known $\Delta_{SW}$. }. 
Equipped with these facts, 
we conclude that computations on both field theory and 
matrix model sides give perfectly the same ${\F}_1$ 
up to some irrelevant constant terms by looking at 
\eqref{a1}, \eqref{a2}, \eqref{a3} and \eqref{a4}.

\section{Summary}
 
We have provided further evidence on the equivalence between a 
recently proposed Dijkgraaf-Vafa matrix model and low-energy dynamics of $\N=2$ asymptotically free $SU(2)$ Yang-Mills theory with 
$N_f=2$, $3$ at the level of ${\F}_1$. 
We utilized the matrix model technique 
which prescribes an universal form of ${\F}_1$. 
Ingredients for computing the 
asymptotically free ${\F}_1$ can be gathered just from 
a classical spectral curve found in 
\cite{Eguchi:2009gf} by decoupling 
very massive flavors from an $N_f=4$ one. 
Showing perfect agreements 
with the field theoretical result, we thus 
extend the equivalence of $Z_{\mathrm{ DV }}$ and 
$Z_{\mathrm{ Nekrasov }}$ at 
next-to-leading order non-trivially.

It will also be interesting to examine whether this check 
gets possible in the superconformal $N_f=4$ case. As shown by Eguchi and Maruyoshi in this situation $d a/d u = K(q_{UV})$, so it is quite tempting to consider 
relations between $q_{UV}$ and the cross-ratio $\ell^2$ of four branch points given a complete 
elliptic integral of the first kind in the 
universal expression of ${\F}_1$ in \eqref{F11}.

\section*{Acknowledgements} 
TST is grateful to Takeo Inami, 
Takahiro Kubota, Wen-Yu Wen and especially 
Kazutoshi Ohta for directing his attention to this field. 
He also thanks Robbert Dijkgraaf 
for an enlightening talk at RIKEN and Kazuhiro Sakai for valuable comments. 
MF would like to thank Masafumi Fukuma 
and Shinji Shimasaki for helpful discussions. 
TST is supported in part by the postdoctoral program at RIKEN. 
YH is supported in part by JSPS research 
fellowships for young scientists.

\end{document}